# High-Accuracy Facial Depth Models derived from 3D Synthetic Data

Faisal Khan
*College of Engineering and Informatics*
*National University Ireland Galway*
Galway, Ireland
f.khan4@nuigalway.ie

Shubhajit Basak
*College of Engineering and Informatics*
*National University Ireland Galway*
Galway, Ireland
s.basak1@nuigalway.ie

Hossein Javidnia
*ADAPT Center, O Reilly Institute*
*Trinity College Dublin Ireland*
*25 Westland Row, Dublin, 2, Ireland*
hossein.javidnia@adaptcentre.ie

Michael Schukat
*College of Engineering and Informatics*
*National University Ireland Galway*
Galway, Ireland
micheal.schukat@nuigalway.ie

Peter Corcoran
*College of Engineering and Informatics*
*National University Ireland Galway*
Galway, Ireland
peter.corcoran@nuigalway.ie

*Abstract*—In this paper, we explore how synthetically generated 3D face models can be used to construct a high-accuracy ground truth for depth. This allows us to train the Convolutional Neural Networks (CNN) to solve facial depth estimation problems. These models provide sophisticated controls over image variations including pose, illumination, facial expressions and camera position. 2D training samples can be rendered from these models, typically in RGB format, together with depth information. Using synthetic facial animations, a dynamic facial expression or facial action data can be rendered for a sequence of image frames together with ground truth depth and additional metadata such as head pose, light direction, etc. The synthetic data is used to train a CNN-based facial depth estimation system which is validated on both synthetic and real images. Potential fields of application include 3D reconstruction, driver monitoring systems, robotic vision systems, and advanced scene understanding.

## I. INTRODUCTION

Estimating human shape, pose, motion and depth from images are fundamental challenges for many multimedia applications and provide information that can be leveraged to enhance quality and immersion in advanced consumer use cases. Examples include scene analysis & understanding, human behavior analysis, driver monitoring for semi-autonomous driving, augmented reality systems and facial expression analysis and facial authentication. Today, state-of-art systems for these use cases will rely on highly optimized convolutional neural networks designed to run on low-power embedded hardware. Such solutions require large, high-quality training datasets.

Facial images, in particular, are at the core of many consumer multimedia systems. They exhibit rich variations in pose, hairstyle, expression, structure and their 2D appearance is affected by external factors such as lighting and camera location. Many face variations can be synthesized using existing advanced 3D tools such as iClone [1] and Blender [2]. Using these tools, it is feasible to generate a large number of synthetic images required for training Convolutional Neural Network (CNN) models. Rendering synthetic facial images would be highly useful for numerous tasks as it can provide enough realism to create various ground truth in terms of occlusions, depth, motion, body-part segmentation, camera and light direction.

The current generation of deep learning models requires the datasets to contain various information and accurate data for the training and evaluation process. The existing human facial datasets do not have the accurate depth information that defines the actual position of each facial element. The depth information in these datasets requires the manual description of the scene, which is an error-prone and time-consuming task especially dealing with video [3]. In such type of facial dataset, they are not sufficiently large and varied enough to learn the CNN models, as a consequence they come with a low performance which restricts real-world applications [4-5].

Recently deep learning-based methodologies have significantly improved the performances of face recognition systems, Human-Computer Interaction (HCI), understanding of 3D scenes for autonomous driving and robotics. An accurate determination of depth within the 3D scene is an important element of these computer vision systems. New emerging applications such as 3D reconstruction, Driver Monitoring Systems (DMS), robotic vision systems for personal robots and advanced HCI modalities require further improvements in short-range depth analysis to better understand and engage with humans.

In this work, we present a method for generating advanced facial models with synthetic data. A method is proposed to generate facial depth information using 3D virtual human and iClone [1] character modelling software. The proposed method can be scaled to produce any number of synthetic facial data by controlling the face animations, scene and camera position.

The main contribution of this research is focused on facial image rendering with the corresponding ground truth depth information. Using the synthetically generated data, we can train CNNs to address the facial depth estimation problem. This approach can enrich the real-world facial datasets required for portrait depth estimation problem.

The rest of the paper is structured as follows: Section II discusses related work and Section III presents the facial models. The application of synthetic facial depth (evaluation) is studied in Section IV. Conclusion and further cautions are discussed in Section V.

## II. RELATED WORK

Facial depth estimation is considered as one of the challenging issues in computer vision, human-computer interaction and virtual reality. It is used in a wide range of applications which includes controlling 3D avatars, human object detection and human-robot interactions [6-11].

Synthetic human facial data is used frequently to augment real data for pose invariant face recognition. By using the 3D morphable model and Basel face model [13, 14], a pipeline is proposed to create synthetic faces [15]. A synthetic dataset for person identification is studied in [16, 17]. The authors used Blender [2] rendering engine to create different realistic illumination conditions including indoor and outdoor scenes and introduce a novel domain adaptation method that uses the synthetic data. In [13], FaceGen Modeller is used for generating facial ground truth using morphable models.

In [19], a large-scale synthetic dataset called (SURREAL) is introduced where the images are rendered from 3D sequences of MoCap data. In [18], synthetic bodies are obtained by utilizing the SMPL body model [18]. This dataset contains more than 6 million frames with ground truth depth, pose and segmentation masks [19].

Very limited work is done on synthetic facial models to explore the field with the available 3D tools and other commercially available software. In this paper, we proposed a method that generates synthetic facial models with many variations in expressions. By controlling the facial animations, camera positions, light positions, body poses, scene illuminations and other scene parameters, the method can be scaled to generate any number of labeled data samples.

## III. FACIAL DEPTH GENERATOR MODEL

Virtual human models are created using the "*Realistic Human 100*" models in iClone [1] software based on the following steps:

### A. The iClone Character Creation Process

iClone character creator [1] is used to create the initial characters of the virtual human faces. The iClone character creator generates humanoid characters and offers a useful 3D rigging option. The facial animation-ready models can be customized with sculpting and morphs. The template of the "*Realistic Human 100*" models is applied to the base body in the character creator as shown in Fig. 1.

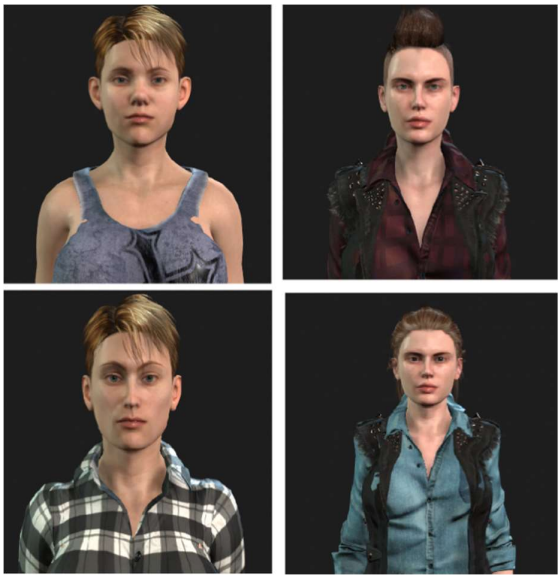

Fig. 1. A sample from the iClone Character creator.

### B. Adding Facial Expressions to Character Models

The virtual human face models are imported from Character creator to iClone [1]. Further, different expressions are added to the face models to introduce variations such as neutral, angry, happy, sad and scared. Fig. 2, show an example of these expressions.

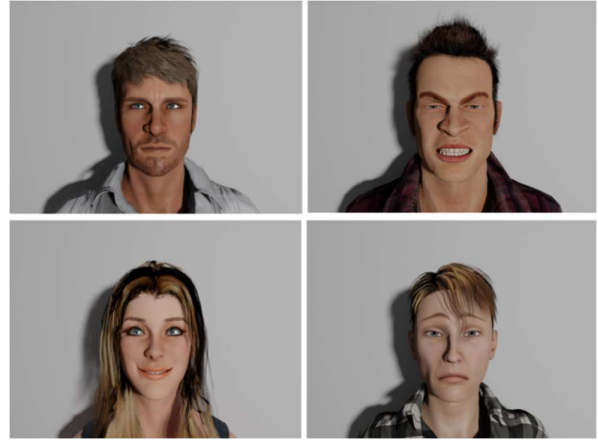

Fig. 2. A sample rendered images of iClone with different expressions (neutral, angry, happy, sad and scared).

### C. Exporting Character Animations to Blender

The created virtual human face models are exported from iClone [1] to Blender [2] in FBX format as it provides appropriate rigging. FBX is a popular 3D file format for exchanging the 3D information as used by many 3D tools including Blender [2]. A sample of an iClone facial model with base body loaded in Blender [2] is shown in Fig. 3.

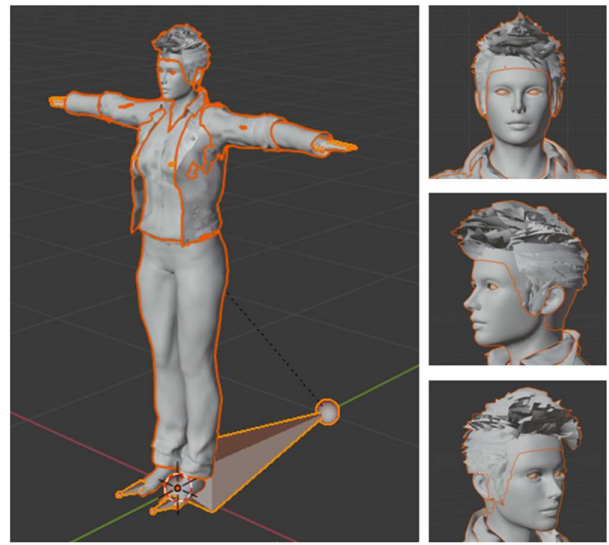

Fig. 3. iClone facial model with base body loaded in Blender.

### D. Rendering 2D Image Data with Ground Truth Depth

In this work, the following steps are taken to obtain the final output. The cameras and lights are placed in a fixed position and the corresponding distance of the models are changed in the range of 700-1000 mm. The focal length and sensor size are set to 60mm and 36mm respectively. The facial models are rotated in the virtual scenes. Fig. 4 shows a sample of the camera and light position with respect to the facial models in Blender [2].

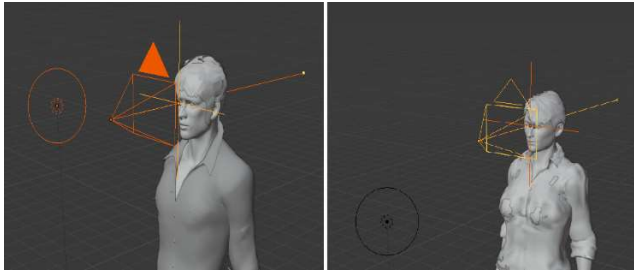

Fig. 4. A sample of the camera and light position with respect to the 3D character.

To generate RGB and depth images of faces in an extensive range of positions, the near and far clip of the camera is set to 0.01 and 5 meters. The facial models are rendered with 480×640 resolution and on a static background image. Fig. 5 shows a few rendered images while the camera position is changed with respect to the facial models.

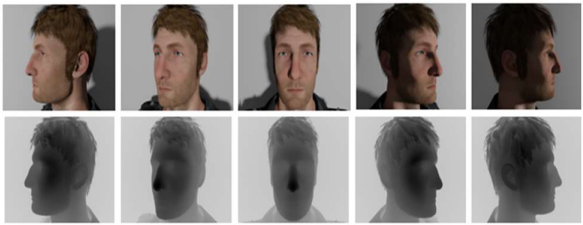

Fig. 5. A facial model with corresponding ground truth depth of a head model from different views.

Fig. 6 illustrates facial models with the corresponding ground truth depth while the camera is positioned at different distances.

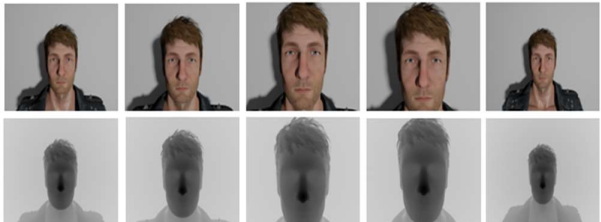

Fig. 6. A facial model with ground truth depth captured at different camera position.

Render passes are set up in Blender [2] to generate the synthetic facial RGB and the corresponding ground truth depth images. To reduce the noise produced during the rendering process, the branched path tracing method is employed. Fig. 7 presents an overview of the noise controlling method in Blender [2].

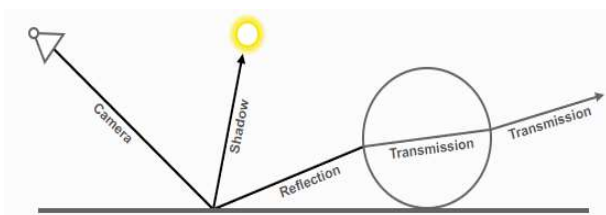

Fig. 7. An overview of the noise control system in Blender.

Afterwards, the images are rendered using Cycles engine and in the perspective view to obtain the RGB images with corresponding facial depth. Fig. 8 demonstrates the workflow of the facial depth generation process, camera and light setting.

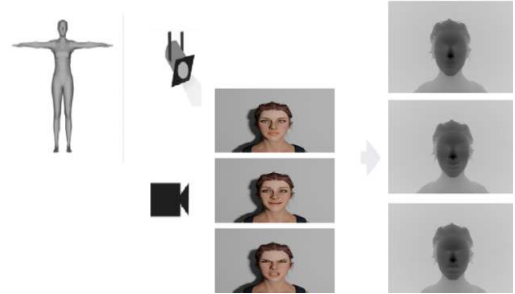

Fig. 8. Rendering configuration in Blender. The left row shows the body shape, light and camera setting; middle row shows the facial RGB and the last row illustrates the corresponding facial depth image.

Fig. 9 shows a few numbers of synthetic male and female models with the corresponding ground truth depth.

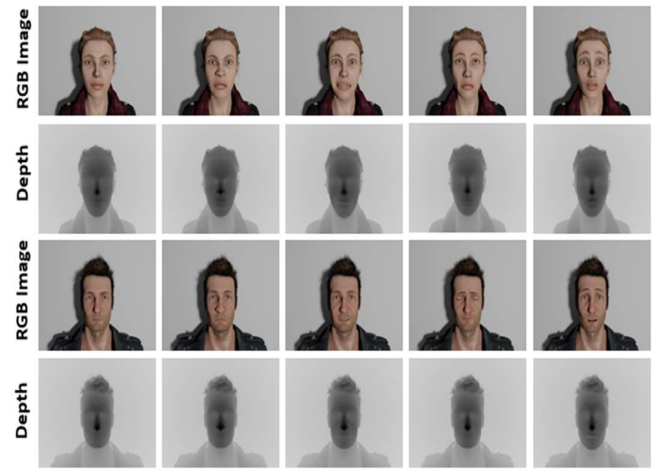

Fig. 9. A sample of the synthetic facial images with different expressions and their corresponding depth maps.

## IV. EVALUATION

In this section, we deliver details about the evaluation of the two-state of the art CNNs on facial depth estimation. The pre-trained monocular depth estimation models DepthDense [19] and MiDas [20] are tested on the rendered synthetic data. Fig. 10, presents a few random synthetic RGB images and the corresponding depth images predicted using DepthDense [19]. Similarly, Fig. 11, shows the synthetic RGB images, predicted depth using MiDas [20] and ground truth images.

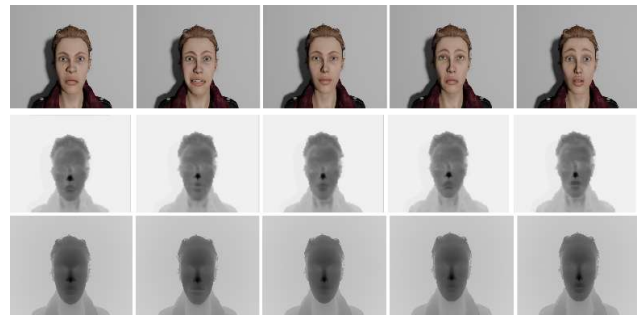

Fig. 10. Sample synthetic RGB images, predicted depth maps by DepthDense [19] and corresponding ground truth.

TABLE I. RESULTS OF THE DEPTHDENSE AND MIDAS MODELS [19, 20].

| No. | Method | Abs Rel | Sq Rel | RMSE | RMSElog | $\delta \prec 1.25$ | $\delta \prec 1.25^2$ | $\delta \prec 1.25^3$ |
|---|---|---|---|---|---|---|---|---|
| 1. | DenseDepth[19] | 0.8765 | 0.7783 | 1.8783 | 0.2260 | 0.2723 | 0.5093 | 0.6912 |
| 2. | MiDas[20] | 0.8876 | 0.9765 | 1.9876 | 0.3323 | 0.3211 | 0.5432 | 0.7635 |

a. Evaluation results of the pre-trained models [19, 20] on the synthetic data.

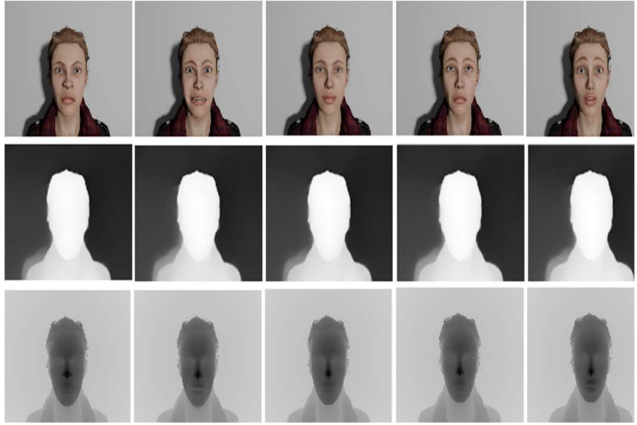

Fig. 11. Sample synthetic RGB images, predicted depth maps by MiDas [20] and corresponding ground truth.

The most common quantitative matrices for evaluating the performance of the pre-trained models including Absolute Relative difference (AbsRel), Root Mean Square Error (RMSE), log Root Mean Square Error (RMSE(log)) and Square Relative error (SqRel) are employed for evaluation purposes. Table 1, demonstrates the evaluation results of the DepthDense and MiDas models [19, 20].

To further evaluate the validity of the synthetic data generated in this paper, we re-trained a few recent CNN-based depth estimation networks [21, 22] on the generated facial data and later fine-tuned the models on real datasets.

Furthermore, we will create additional variations and augmentations in the synthetic facial depth data to grow the final training dataset. It is expected that this will further increase the accuracy of these deep learning-based CNN networks when tested on real data.

A more detailed set of experiments providing a comprehensive evaluation of these depth sensing CNNs will be presented at the ISSC conference.

## V. CONCLUSION AND FUTURE RESEARCH

In this research paper, we proposed an advanced synthetic facial data generation pipeline. The facial images are generated from 3D virtual human models by rendering different variations of face poses, head poses and lighting conditions. Blender [2] rendering engine is used to generate the output as it allows changing different parameters such as lights position, camera parameters and keyframe values.

The proposed framework has a potential to generate a great number of synthetic facial images. The synthetic 3D models can be used in different 3D environments if scaled properly. This will allow simulating real-world scenarios by controlling the camera position, intrinsic parameters and lighting conditions.

The generated dataset can be used for training and validation of deep learning methods with the focus on natural face modelling, portrait 3D reconstruction and beautification.

In our future work, we will explore the potentials of the deep learning methods on direct facial 3D reconstruction using the synthetically generated data.